\documentstyle[preprint,floats,aps,psfig,epsfig]{revtex}
\tightenlines
   \newcommand{\be}{\begin{equation}}
   \newcommand{\ee}{\end{equation}}
   \newcommand{\bea}{\begin{eqnarray}}
   \newcommand{\eea}{\end{eqnarray}}

\begin{document}
\draft
\title{Critical phenomena near the antiferromagnetic quantum critical point of
Heavy-Fermions}
\author{M. Lavagna$^{1,*}$ and C. P\'epin$^2$}
\address{$^1$Commissariat \`a l'Energie Atomique, DRFMC /SPSMS, 
17, rue des Martyrs,
   38054 Grenoble Cedex 9, France}
\address{$^2$University of Oxford, Department of Physics, 1 Kable Road, OX1 3NP Oxford, UK}

\maketitle

\begin{abstract}
We present a study of the critical phenomena around the quantum 
critical point in heavy-fermion systems. In the framework of the 
$S=1/2$ Kondo lattice model, we introduce an extended decoupling
scheme of the Kondo interaction which allows one to treat the spin
fluctuations and the Kondo effect on an equal footing. The calculations,
developed in a self-consistent one-loop approximation, lead to the 
formation of a damped collective mode with a dynamic exponent $z=2$
in the case of an antiferromagnetic instability. The system 
displays a quantum-classical crossover at finite temperature depending
how the energy of the mode, on the scale of the magnetic correlation length,
compares to $k_B T$. The low temperature behavior, in the different regimes 
separated by the crossover temperatures, is then discussed for both 
2- and 3-dimensional systems.
\end{abstract}


\widetext

\widetext
\leftskip10.8pt \rightskip10.8pt

\bigskip

\section{\protect\bigskip Introduction}
\par
A central issue in the thermal properties of heavy-fermion compounds is the discovery of a non-Fermi liquid behavior in systems close to an antiferromagnetic quantum phase transition \cite{santa}. Such a behavior has been discovered in a series of compounds containing cerium or uranium, for example, $CeCu_{6-x}R_{x}$ ($R$=$Au$, $Ag$) \cite{lohneysen}, $CeIn_{3}$, $CePd_{2}Si_{2}$ \cite{mathur}, $CeNi_{2}Ge_{2}$ \cite{steglich}, $U_{1-x}Y_{x}Pd_{3}$ \cite{seaman} and $Ce_{x}La_{1-x}Ru_{2}Si_{2}$ 
\cite{kambe}. In $CeCu_{5.9}Au_{0.1}$, the specific heat $C$ depends
on T as $C/T\sim -ln(T/T_{0})$, the magnetic susceptibility as $\chi \sim
1-\alpha \sqrt{T}$, and the T-dependent part of the resistivity as $\Delta
\rho \sim T$ (instead of $C/T\sim \chi \sim Const$ and $\Delta \rho \sim
T^{2}$ in the Fermi liquid state). Most interestingly, the breakdown of the
Fermi liquid behavior can be tuned by alloying (chemical pressure) or by applying a hydrostatic pressure or a magnetic field. The origin of this NFL behavior is a problem of current debate. 
\par
Different scenarios have been proposed in order to elucidate the non-Fermi liquid (NFL) behavior at the quantum critical point (QCP). On the one hand, there are scenarios involving strong disorder to account for the non-Fermi liquid properties, including (i) a distribution of the Kondo temperature $P(T_{K})$ \cite{miranda} induced by disorder following a distribution of the local density of states, and (ii) the possibility of the formation of a Griffiths phase at the quantum critical point \cite{castroneto,narayanan}. On the other hand, starting from a clean or rather weakly disordered system, there is a scenario (iii) refering to the proximity of a quantum critical point in a theory of itinerant antiferromagnetism 
\cite {andraka,millis93,moriya95,continentino,schlottmann,coleman,si,rosch99}, in which the Fermi surface is coupled to critical collective antiferromagnetic excitations. An alternative possibility has been recently put forward \cite {schroder} on the basis of an analysis of the neutron scattering measurements in $CeCu_{5.9}Au_{0.1}$. The unusual spin dynamics that has been observed suggests (iv) the existence of critically screened local magnetic excitations due to the formation of local moments. Finally, it has been proposed (v) a variant of the quadropular Kondo effect \cite{cox} in which the interactions of the rare earth (actinide) internal degrees of freedom with the conduction electrons are reduced to a single impurity multichannel Kondo effect. As it stands, the situation remains highly controversial.
\par
In this paper, we adopt the point of view (iii) of the proximity of a quantum
critical point. The effect of a non zero temperature on quantum critical
points is a long-standing problem encountered in itinerant magnetism. We
refer to the papers of Hertz \cite{hertz} and Millis \cite
{millis93} using renormalization group techniques,
and of Moriya \cite{moriya} introducing a self-consistent renormalization (SCR) theory of
spin-fluctuations. These theories yield a rich phase diagram with, in
addition to the ordering temperature $T_{N}$, a series of crossover
temperatures separating different regimes of behaviors. They are well
adapted for the treatment of spin fluctuations as present in the single-band
Hubbard model but essentially do not account for the Kondo effect. 
\par
Indeed,
the model which is believed to describe the physics of heavy fermions is the
Kondo lattice model (KLM) \cite{doniach} derived from the periodic Anderson model (PAM) in
a given limit. The large $N$ expansions \cite {millis87} which have been carried out for these
models (where $N$ represents simultaneously the degeneracy of the conduction
electrons and of the spin channels) give a good description of the Kondo
effect. But, unlike the above-mentioned spin-fluctuation theories, they fail
to account for the spin-fluctuations since the RKKY interactions only appear
at the order $1/N^{2}$ \cite{houghton}. 
\par
Hence our motivation to set up an approach of the $S=1/2$ Kondo lattice model ($N=2$)\ that enlarges the standard $1/N$\
expansion theories up on the spin-fluctuation effects. Our previous paper 
\cite{pepin}, presented such an approach by performing a generalized Hubbard-Stratonovich
transformation on the Kondo interaction term of the Hamiltonian which makes
three types of fields (magnetization densities and Kondo-type) appear on an equal footing. The
dynamical spin susceptibility was derived in a one-loop expansion associated
with the gaussian fluctuations of the magnetization density fields around their
saddle-point values. 
\par
With the aim to describe the critical phenomena around the quantum
critical point of the heavy-fermion systems, we propose to push the treatment of the $S=1/2$ Kondo lattice model in a
self-consistent one-loop approximation in complete analogy with the SCR
theory of spin fluctuations developed for the Hubbard model \cite{moriya,nozieres,klenin}. As in itinerant
magnetism, we will show how the system displays a quantum-classical crossover
at finite temperature depending whether the temperature is lower or higher
than the characteristic energy-scale of the collective mode associated with
the magnetic instability. The low temperature behavior in the different
regimes essentially depends on the value of the dynamic exponent $z$ of the
mode and on the dimensionality $d$ of the problem. In the case of an antiferromagnetic
instability, $z$ is found to be equal to $2$. Heavy-fermion systems are
usually believed to be three-dimensional. However, a recent proposal \cite{rosch} based
on the neutron scattering data obtained in $CeCu_{6-x}Au_{x}$ stipulates
that the critical magnetic fluctuations which the quasiparticles are coupled
to, are simply two-dimensional. So both situations $d=2$ and $d=3$ will be
considered.
\par
The rest of the paper is organized as follows. In part II, we introduce the
self-consistent one-loop approximation method for the Kondo lattice model
using a generalized Hubbard-Stratonovich transformation on the Kondo
interaction term. The approach combines both aspects of the $1/N$ expansion
and of the self-consistent renormalization theory of spin fluctuations. The
calculations, presented in the functional integral formalism, closely follow
the presentation made by Hertz and Klenin \cite{klenin} for the Hubbard model. In part
III, we extract the general expression of the dynamical spin susceptibility
extending the results of the previous paper \cite{pepin} to a self-consistent treatment.
Part IV gives a discussion on the nature of the damped collective mode
associated to the proximity of the magnetic transition
with a dynamic exponent $z=2$ in the antiferromagnetic case. In part V, we
derive the resulting quantum-classical crossover. This occurs at a finite
temperature T, which depends on how the energy of the mode, on the scale of the magnetic 
correlation length $\xi $
compares to $k_{B}T$. The low temperature behavior of the
system in the different regimes, separated by the crossover temperatures, is
then discussed in part VI. Concluding remarks follow in section VII.

\section{Self-consistent one-loop approximation to the Kondo lattice model}
\par
We consider the Kondo lattice model (KLM) as given by a periodic array of
Kondo impurities with an average number of conduction electrons per site $
n_{c}<1$. In the grand canonical ensemble, the Hamiltonian is written as

\begin{equation}
\label{1}
H=\sum_{k\sigma }\varepsilon _{k}c_{k\sigma }^{\dagger }c_{k\sigma
}+J\sum\limits_{i}{\bf {S}}_{fi}\cdot \sum\limits_{\sigma \sigma ^{\prime
}}c_{i\sigma }^{\dagger }\mbox{\boldmath{$\tau$}}_{\sigma \sigma ^{\prime
}}c_{i\sigma ^{\prime }}-\mu N_{S}(\frac{1}{N_{S}}\sum_{k\sigma }c_{k\sigma
}^{\dagger }c_{k\sigma }-n_{c}),
\end{equation}
in which ${{\bf {S}}_{fi}}$ represents the spin ($S=1/2$) of the impurities
distributed on the sites (in number $N_{S}$) of a periodic lattice; $
c_{k\sigma }^{\dagger }$ is the creation operator of the conduction electron
of momentum ${\bf {k}}$ and spin quantum number $\sigma $ characterized by
the energy $\epsilon _{k}=-\sum\limits_{<i,j>}t_{ij}\exp \left( i{\bf {k}}.
{\bf {R}}_{ij}\right) $ and the chemical potential $\mu $; $
\mbox{\boldmath{$\tau$}}$ are the Pauli matrices $\left( \mbox{\boldmath{$
\tau$}}^{x},\mbox{\boldmath{$\tau$}}^{y},\mbox{\boldmath{$\tau$}}^{z}\right) 
$ and $\mbox{\boldmath{$\tau$}}^{0}$ the unit matrix; J is the
antiferromagnetic Kondo interaction $\left( J>0\right) $.

\bigskip

We use the Abrikosov pseudofermion representation of the spin ${\bf {S}}
_{fi}=\sum\limits_{\sigma \sigma ^{\prime }}f_{i\sigma }^{\dagger }
\mbox{\boldmath{$\tau$}}_{\sigma \sigma ^{\prime }}f_{i\sigma ^{\prime }}$.
The projection into the physical subspace is implemented by a local
constraint

\begin{equation}
\label{2}
Q_{i}=\sum\limits_{\sigma }f_{i\sigma }^{+}f_{i\sigma }-1=0.
\end{equation}
A Lagrange multiplier $\lambda _{i}$ is introduced to enforce the local
constraint $Q_{i}$. Since $[Q_{i},H]=0$, $\lambda _{i}$ is time-independent.
In this representation, the partition function of the KLM can be expressed
as a functional integral over the coherent states of the fermion fields

\begin{equation}
\label{3}
Z=\int {\cal D}c_{i\sigma }{\cal D}f_{i\sigma }d\lambda _{i}\exp \left[
-\int_{0}^{\beta }{\cal L}(\tau )d\tau \right] ,
\end{equation}
where the Lagrangian ${\cal L}(\tau )$ is given by

\[
{\cal L}(\tau )={\cal L}_{0}(\tau )+H_{0}(\tau )+H_{J}(\tau ) 
\]

\[
{\cal L}_{0}(\tau )=\sum_{i\sigma }c_{i\sigma }^{\dagger}\partial _{\tau
}c_{i\sigma }+f_{i\sigma }^{\dagger}\partial _{\tau }f_{i\sigma } 
\]

\[
H_{0}(\tau )=\sum_{k\sigma }\epsilon _{k}c_{k\sigma }^{\dagger}c_{k\sigma
}-\mu N_{S}\left( \frac{1}{N_{S}}\sum_{k\sigma }c_{k\sigma
}^{\dagger}c_{k\sigma }-n_{c}\right) +\sum_{i}\lambda _{i}Q_i 
\]

\[
H_{J}(\tau )=J\sum_{i}{\bf {S}}_{fi}\cdot {\bf {S}}_{ci}, 
\]
with ${\bf {S}}_{c_{i}}=\sum\limits_{\sigma \sigma ^{\prime }}c_{i\sigma
}^{\dagger }\mbox{\boldmath{$\tau$}}_{\sigma \sigma ^{\prime }}c_{i\sigma
^{\prime }}$.

Following \cite{pepin}, we perform a Hubbard-Stratonovich
transformation on the Kondo interaction term $H_{J}(\tau )$ which makes the
fields $\Phi _{i}$, $\Phi _{i}^{\ast }$ and $
\mbox{\boldmath{$\xi$}}_{f_{i}},$ $\mbox{\boldmath{$\xi$}}_{c_{i}}$ appear.
We get

\begin{equation}
\label{4}
Z=\int d\Phi _{i}d\Phi _{i}^{\ast }d\mbox{\boldmath{$\xi$}}_{f_{i}}d
\mbox{\boldmath{$\xi$}}_{c_{i}}Z(\Phi _{i},\Phi _{i}^{\ast },
\mbox{\boldmath{$\xi$}}_{f_{i}},\mbox{\boldmath{$\xi$}}_{c_{i}}),
\end{equation}
with 
\[
Z(\Phi _{i},\Phi _{i}^{\ast },\mbox{\boldmath{$\xi$}}_{f_{i}},
\mbox{\boldmath{$\xi$}}_{c_{i}})=\int {\cal D}c_{i\sigma }{\cal D}f_{i\sigma
}d\lambda _{i}\exp \left[ -\int_{0}^{\beta }{\cal L}^{\prime }(\tau )d\tau 
\right] 
\]
\[
{\cal L}^{\prime }(\tau )={\cal L}_{0}(\tau )+H_{0}(\tau )+H_{J}^{\prime
}(\tau ) 
\]

\[
H_{J}^{\prime }(\tau )=\sum_{i\sigma \sigma ^{\prime }}\left( 
\begin{array}{cc}
c_{i\sigma }^{\dagger } & f_{i\sigma }^{\dagger }
\end{array}
\right) \left( 
\begin{array}{cc}
-J_{S}i\mbox{\boldmath{$\xi$}}_{f_{i}}\cdot \mbox{\boldmath{$\tau$}}_{\sigma
\sigma ^{\prime }} & J_{C}\Phi _{i}^{\ast }\tau _{\sigma \sigma ^{\prime
}}^{0} \\ 
J_{C}\Phi _{i}\tau _{\sigma \sigma ^{\prime }}^{0} & -J_{S}i
\mbox{\boldmath{$\xi$}}_{c_{i}}\cdot \mbox{\boldmath{$\tau$}}_{\sigma \sigma
^{\prime }}
\end{array}
\right) \left( 
\begin{array}{c}
c_{i\sigma ^{\prime }} \\ 
f_{i\sigma ^{\prime }}
\end{array}
\right) +J_{C}\sum_{i}\Phi _{i}^{\ast }\Phi _{i}+J_{S}\sum_{i}
\mbox{\boldmath{$\xi$}}_{f_{i}}.\mbox{\boldmath{$\xi$}}_{c_{i}}, 
\]
with $J_{S}=J/4$ and $J_{C}=J/3$.

The saddle-point solution is obtained when neglecting the space and time 
dependence of the different fields. Details of the solution are given in 
Appendix A. It gives rise to the band structure illustrated in Figure 1 
with the formation of a Kondo or Abrikosov-Suhl resonance pinned at the 
Fermi level and split by a hybridization gap defining two bands $\alpha$ 
and $\beta$. The saddle-point solution transposes to N=2 the large-N 
results obtained within the slave-boson mean-field theories (SBMFT).

Then we consider the self-consistent one-loop approximation beyond the
saddle-point solution in the magnetically-disordered phase $
\mbox{\boldmath{$\xi$}}_{f_{0}}=\mbox{\boldmath{$\xi$}}_{c_{0}}=0$.
Following Read and Newns \cite{read}, we take advantage of the local U(1)
gauge transformation of the lagrangian ${\cal L}^{\prime }(\tau )$: 
$\Phi_{i}\rightarrow r_{i}\exp (i\theta _{i})$; $f_{i}\rightarrow f_{i}^{\prime
}\exp (i\theta _{i})$; $\lambda _{i}\rightarrow \lambda _{i}^{^{\prime
}}+i~\partial \theta _{i}/\partial \tau $. We use the radial gauge in which
the modulus of both fields $\Phi _{i}$ and $\Phi _{i}^{\ast }$ are the
radial field $r_{i}$, and their phase $\theta _{i}$ (via its time
derivative) is incorporated into the Lagrange multiplier $\lambda _{i}$
which turns out to be a field. Use of the radial instead of the cartesian
gauge bypasses the familiar complications of infrared divergences associated
with unphysical Goldstone bosons. We let the fields fluctuate away from
their saddle-point values : $r_{i}=r_{0}+\delta r_{i}$, $\lambda
_{i}=\lambda _{0}+\delta \lambda _{i}$, ${\bf {\xi }_{f_{i}}=\delta 
\mbox{\boldmath{$\xi$}}_{f_{i}}}$ and ${\bf {\xi }_{c_{i}}=\delta 
\mbox{\boldmath{$\xi$}}_{c_{i}}}$. After integrating out the Grassmann
variables in the partition function in Equation (\ref{3}), we express $Z$
as a functional integral over the bosonic fields $(r_{i},\lambda _{i},
\mbox{\boldmath{$\xi$}}_{f_{i}},\mbox{\boldmath{$\xi$}}_{c_{i}})$

\begin{equation}
\label{5}
Z=\int {\cal D}r_{i}{\cal D}\lambda _{i}{\cal D}\mbox{\boldmath{$\xi$}}
_{f_{i}}{\cal D}\mbox{\boldmath{$\xi$}}_{c_{i}}\exp [-S_{eff}(r_{i},\lambda
_{i},\mbox{\boldmath{$\xi$}}_{f_{i}},\mbox{\boldmath{$\xi$}}_{c_{i}})],
\end{equation}
where the effective action is

\begin{eqnarray}
S_{eff}(r,\lambda ,\mbox{\boldmath{$\xi$}}_{f},\mbox{\boldmath{$\xi$}}
_{c}) &=&-\sum_{k,i\omega _{n}}Ln~Det{\bf G}^{-1}({\bf {k}},i\omega
_{n},r_{i},\lambda _{i},\mbox{\boldmath{$\xi$}}_{f_{i}},\mbox{\boldmath{$
\xi$}}_{c_{i}}) \nonumber \\
&+&\beta ~[~J_{C}\sum_{i}r_{i}^{2}+J_{S}\sum_{i}
\mbox{\boldmath{$\xi$}}_{f_{i}}\cdot \mbox{\boldmath{$\xi$}}
_{c_{i}}+N_{S}(\mu n_{c}-\lambda _{0})], 
\end{eqnarray}
with : 
\begin{eqnarray}
&\left[ {\bf G}^{-1}(i\omega _{n},r,\lambda ,\mbox{\boldmath{$\xi$}}_{f},
\mbox{\boldmath{$\xi$}}_{c})\right] _{ij}^{\sigma \sigma ^{\prime }}=& 
\nonumber \\
&\left( 
\begin{array}{cc}
\lbrack (-i\omega _{n}-\mu )\delta _{ij}-t_{ij}]\delta _{\sigma \sigma
^{\prime }}-J_{S}i\mbox{\boldmath{$\xi$}}_{f_{i}}.\mbox{\boldmath{$\tau$}}
_{\sigma \sigma ^{\prime }}\delta _{ij} & (\sigma _{0}+J_{C}\delta
r_{i})\delta _{\sigma \sigma ^{\prime }}\delta _{ij} \\ 
(\sigma _{0}+J_{C}\delta r_{i})\delta _{\sigma \sigma ^{\prime }}\delta _{ij}
& [-i\omega _{n}+\varepsilon _{f}+\delta \lambda _{i}]\delta _{\sigma \sigma
^{\prime }}\delta _{ij}-J_{S}i\mbox{\boldmath{$\xi$}}_{c_{i}}.
\mbox{\boldmath
{$\tau$}}_{\sigma \sigma ^{\prime }}\delta _{ij}
\end{array}
\right).&
\end{eqnarray}

As in the conventional approaches to critical phenomena, we have
integrated out the fermions and thereby reduced the problem to the study of
an effective bosonic theory describing fluctuations of the ordering fields.
But contrary to Hertz \cite{hertz} and Millis \cite{millis93}, who considered 
a single field corresponding to the magnetization density of one type of electron, 
our approach, starting from a different microscopic model, 
extends the description to the case of several fields i.e. the Kondo fields 
$(r_{i},\lambda _{i})$ on the one hand, and the $f$- and $c$- magnetization
densities on the other.

We then propose to solve this field theory in a variational way. We replace
the full expression $S_{eff}(r_{i},\lambda _{i},\mbox{\boldmath{$\xi$}}
_{f_{i}},\mbox{\boldmath{$\xi$}}_{c_{i}})-S_{eff}^{(0)}$ (where $
S_{eff}^{(0)}$ and $F^{(0)}$ are the zeroth order effective action and the
related free energy) by a trial quadratic form $\Delta \widetilde{S}
_{eff}^{(2)}(r,\lambda ,\mbox{\boldmath{$\xi$}}_{f},\mbox{\boldmath{$\xi$}}
_{c})=S_{eff}^{(2)}(r_{i},\lambda _{i},\mbox{\boldmath{$\xi$}}_{f_{i}},
\mbox{\boldmath{$\xi$}}_{c_{i}})-S_{eff}^{(0)}$ in the fields $\delta r_{i}$
, $\delta \lambda _{i}$, $\delta \mbox{\boldmath{$\xi$}}_{f_{i}}$ and $\delta 
\mbox{\boldmath{$\xi$}}_{c_{i}}$

\begin{eqnarray}
\label{6}
\Delta \widetilde{S}_{eff}^{(2)}(r,\lambda ,\mbox{\boldmath{$\xi$}}_{f},
\mbox{\boldmath{$\xi$}}_{c}) &=&\frac{1}{\beta }\sum_{{\bf {q}},i\omega
_{\nu }}[\left( 
\begin{array}{cc}
\delta r_{q,\nu } & \delta \lambda _{q,\nu }
\end{array}
\right) \widetilde{{\bf D}}_{C}^{-1}({\bf {q}},i\omega _{\nu })\left( 
\begin{array}{c}
\delta r_{-q,-\nu } \\ 
\delta \lambda _{-q,-\nu }
\end{array}
\right) \nonumber \\
&+&\left( 
\begin{array}{cc}
\delta \xi _{f\,q,\nu }^{z} & \delta \xi _{cq,\nu }^{z}
\end{array}
\right) \widetilde{{\bf D}}_{S}^{\Vert -1}({\bf {q}},i\omega _{\nu })\left( 
\begin{array}{c}
\delta \$xi _{f-q,-\nu }^{z} \\ 
\delta \xi _{c-q,-\nu }^{z}
\end{array}
\right)  \nonumber \\
&+&\left( 
\begin{array}{cc}
\delta \xi _{fq,\nu }^{+} & \delta \xi _{cq,\nu }^{+}
\end{array}
\right) \widetilde{{\bf D}}_{S}^{\bot -1}({\bf {q}},i\omega _{\nu })\left( 
\begin{array}{c}
\delta \xi _{f-q,-\nu }^{-} \\ 
\delta \xi _{c-q,-\nu }^{-}
\end{array}
\right) \nonumber \\
&+&\left( 
\begin{array}{cc}
\delta \xi _{fq,\nu }^{-} & \delta \xi _{cq,\nu }^{-}
\end{array}
\right) \widetilde{{\bf D}}_{S}^{\bot -1}({\bf {q}},i\omega _{\nu })\left( 
\begin{array}{c}
\delta \xi _{f-q,-\nu }^{+} \\ 
\delta \xi _{c-q,-\nu }^{+}
\end{array}
\right) ].
\end{eqnarray}
where the indices $q,\nu$ stand for $({\bf {q}},i\omega
_{\nu })$. The elements of the matrices $\widetilde{{\bf D}}_{C}^{-1}({\bf {q}},i\omega
_{\nu })$, $\widetilde{{\bf D}}_{S}^{\Vert -1}({\bf {q}},i\omega _{\nu })$ and $
\widetilde{{\bf D}}_{S}^{\bot -1}({\bf {q}},i\omega _{\nu })$ are determined
variationally by minimizing the free energy or, more precisely, the upper
bound to the free energy according to the Feynman variational principle $
F\leq \widetilde{F}_{eff}^{(2)}+1/\beta \left\langle S_{eff}(r,\lambda ,
\mbox{\boldmath{$\xi$}}_{f},\mbox{\boldmath{$\xi$}}_{c})-\widetilde{S}
_{eff}^{(2)}(r,\lambda ,\mbox{\boldmath{$\xi$}}_{f},\mbox{\boldmath{$\xi$}}
_{c})\right\rangle $ where $\widetilde{F}_{eff}^{(2)}$ represents the free
energy related to $\widetilde{S}_{eff}^{(2)}(r,\lambda ,\mbox{\boldmath{$
\xi$}}_{f},\mbox{\boldmath{$\xi$}}_{c})$ and $\left\langle A\right\rangle $
means the average of the quantity $A(r_{i},\lambda _{i},\mbox{\boldmath{$
\xi$}}_{f_{i}},\mbox{\boldmath{$\xi$}}_{c_{i}})$ over the distribution of
fields $\exp [-\Delta \widetilde{S}_{eff}^{(2)}(r,\lambda ,
\mbox{\boldmath{$\xi$}}_{f},\mbox{\boldmath{$\xi$}}_{c})]$ i.e. 
\[
\left\langle A\right\rangle =\frac{(\int {\cal D}r_{i}{\cal D}\lambda _{i}
{\cal D}\mbox{\boldmath{$\xi$}}_{f_{i}}{\cal D}\mbox{\boldmath{$\xi$}}
_{c_{i}}A(r_{i},\lambda _{i},\mbox{\boldmath{$\xi$}}_{f_{i}},
\mbox{\boldmath{$\xi$}}_{c_{i}})\exp [-\Delta \widetilde{S}
_{eff}^{(2)}(r,\lambda ,\mbox{\boldmath{$\xi$}}_{f},\mbox{\boldmath{$\xi$}}
_{c})])}{(\int {\cal D}r_{i}{\cal D}\lambda _{i}{\cal D}\mbox{\boldmath{$
\xi$}}_{f_{i}}{\cal D}\mbox{\boldmath{$\xi$}}_{c_{i}})\exp [-\Delta 
\widetilde{S}_{eff}^{(2)}(r,\lambda ,\mbox{\boldmath{$\xi$}}_{f},
\mbox{\boldmath{$\xi$}}_{c})])}.
\]

After minimization, one can show that $[{\bf D}_{C}^{-1}({\bf {q}},i\omega
_{\nu })]_{r\lambda }$ is equal to $(1/2)\partial ^{2}S_{eff}(r_{i},\lambda
_{i},\mbox{\boldmath{$\xi$}}_{f_{i}},\mbox{\boldmath{$\xi$}}
_{c_{i}})/(\partial r_{q,\nu }\partial \lambda _{-q,-\nu })$ averaged over
the mentioned distribution of fields (and equivalent expressions for the
other elements). Explicitly we obtain,

\begin{equation}
\label{7}
\widetilde{{\bf D}}_{C}^{-1}({\bf {q}},i\omega _{\nu })=\left( 
\begin{array}{cc}
J_{C}[1-J_{C}(\left\langle 
\overline{\varphi }_{2}({\bf {q}},i\omega _{\nu })\right\rangle
+\left\langle \overline{\varphi }_{m}({\bf {q}},i\omega _{\nu })\right\rangle)]
& -J_{C}\left\langle \overline{\varphi }_{1}({\bf {q}},i\omega _{\nu
})\right\rangle \\ 
-J_{C}\left\langle \overline{\varphi }_{1}({\bf {q}},i\omega _{\nu
})\right\rangle & -\left\langle \overline{\varphi }_{ff}({\bf {q}},i\omega
_{\nu })\right\rangle
\end{array}
\right)
\end{equation}
\[
\widetilde{{\bf D}}_{S}^{\Vert -1}({\bf {q}},i\omega _{\nu })=\left( 
\begin{array}{cc}
J_{S}^{2}\left\langle \varphi _{cc}^{\Vert }({\bf {q}},i\omega _{\nu
})\right\rangle & J_{S}[1+J_{S}\left\langle \varphi _{fc}^{\Vert }({\bf {q}}
,i\omega _{\nu })\right\rangle ] \\ 
J_{S}[1+J_{S}\left\langle \varphi _{cf}^{\Vert }({\bf {q}},i\omega _{\nu
})\right\rangle ] & J_{S}^{2}\left\langle \varphi _{ff}^{\Vert }({\bf {q}}
,i\omega _{\nu })\right\rangle
\end{array}
\right) , 
\]
and equivalent expression for the transverse spin part $\widetilde{{\bf D}}
_{S}^{\bot -1}({\bf {q}},i\omega _{\nu })$. The different bubbles $
\left\langle \varphi ({\bf {q}},i\omega _{\nu })\right\rangle $ in the
latter expressions represent the susceptibilities $\varphi ({\bf {q}}
,i\omega _{\nu })$ (cf. appendix B) averaged over the distribution of fields 
$\exp [-\Delta \widetilde{S}_{eff}^{(2)}(r,\lambda ,\mbox{\boldmath{$\xi$}}
_{f},\mbox{\boldmath{$\xi$}}_{c})]$. At that point, the problem is highly
complex and cannot be solved exactly. Therefore, we introduce a local and
instantaneous approximation which is equivalent to averaging over gaussian
distributions of local fields $(r_{i},\lambda _{i},\mbox{\boldmath{$\xi$}}
_{f_{i}},\mbox{\boldmath{$\xi$}}_{c_{i}})$ with variances $\sigma _{r}$, $
\sigma _{\lambda }$, $\sigma _{f}$ and $\sigma _{c}$ defined as

\begin{equation}
\label{8}
\sigma _{f}^{2}=3\sum_{{\bf {q}},i\omega _{\nu }}[{\widetilde{{\bf D}}
_{S}^{\Vert }({\bf {q}},i\omega _{\nu })}_{ff}],
\end{equation}
and similar expressions for the three other variances. Eq.(\ref{8})
is obtained from Eq.(\ref{6}) by letting the summation over $(q,i\omega _{\nu })$
act on the matrix elements only, hence defining the local and instantaneous
fluctuations of the various degrees of freedom. One expects that at low
temperatures, the fluctuations of $r_{i}$, $\lambda _{i}$ and $
\mbox{\boldmath{$\xi$}}_{c_{i}}$ only bring $T^{2}$ corrections to their
saddle-point values. Those corrections are negligible compared to the $T$-dependence 
brought by the fluctuations of $\mbox{\boldmath{$\xi$}}_{f_{i}}$
and will not be considered here. So we will neglect in the following the
fluctuations of $r_{i}$, $\lambda _{i}$ and $\mbox{\boldmath{$\xi$}}_{c_{i}}$
and focus on the fluctuations of $\mbox{\boldmath{$\xi$}}_{f_{i}}$ only.
Then $\left\langle \varphi ({\bf {q}},i\omega _{\nu })\right\rangle $ is
simply given by

\begin{equation}
\label{9}
\left\langle \varphi ({\bf {q}},i\omega _{\nu })\right\rangle =\frac{(\int {\cal D}
\xi _{f_{i}}\varphi ({\bf {q}},i\omega _{\nu },r_{0},\lambda _{0},\xi
_{f_{i}},\xi _{c_{0}})\exp [-\xi _{f_{i}}^{2}/(2\sigma _{f}^{2})])}{(\int 
{\cal D}\xi _{f_{i}}\exp [-\xi _{f_{i}}^{2}/(2\sigma _{f}^{2})])}.
\end{equation}

Eqs. (\ref{8}-\ref{9}) are the basic equations of the paper from which the dynamical spin
susceptibility is extracted and the critical phenomena around the quantum
critical point (QCP) are studied. Let us note that, when the averages
acting on the various bubbles $\varphi ({\bf {q}},i\omega _{\nu })$ in Eq.(\ref{7})
are omitted, one recovers the standard results of the random phase
approximation. As pointed out by Lonzarich and Taillefer \cite{lonzarich}, the corresponding
self-consistent renormalized spin fluctuation SCR-SF procedure can be
analyzed within the Ginzburg-Landau formalism in which the local free energy
is expanded in terms of a small and slowly varying varying order parameter $
{\bf m}({\bf r})$ as $f({\bf r})=f_{0}+\frac{1}{2}a\left| {\bf m}({\bf r}
)\right| ^{2}+\frac{1}{4}b\left| {\bf m}({\bf r})\right| ^{4}+\frac{1}{2}
c\sum\limits_{i}\left| {\bf \nabla }m^{i}({\bf r})\right| ^{2}+O(6)$ and the
following approximation is made $\left| {\bf m}({\bf r})\right|
^{4}=\left\langle \left| {\bf m}({\bf r})\right| ^{2}\right\rangle \left| 
{\bf m}({\bf r})\right| ^{2}$. Hence $\left\langle \left| {\bf m}({\bf r}
)\right| ^{2}\right\rangle $ is evaluated via the
fluctuation-dissipation theorem.

\section{\protect\bigskip Dynamical spin susceptibility}

Let us now derive the expression of the dynamical spin susceptibility. For
that purpose, we study the linear response $M_{f}$ to the source-term $-2
{\bf {S}}_{f}.{\bf {B}}$ (we consider ${\bf {B}}$ colinear to the ${\bf {z}}$
-axis). The effect on the partition function expressed in Equation (\ref{4}
) is to change the $H_{J}^{\prime }(\tau )$ to $H_{J}^{\prime
B}(\tau )$ 
\begin{eqnarray}
\label{10}
H_{J}^{\prime B}(\tau )=&\sum_{i\sigma \sigma ^{\prime }}\left( 
\begin{array}{cc}
c_{i\sigma }^{\dagger } & f_{i\sigma }^{\dagger }
\end{array}
\right) \left( 
\begin{array}{cc}
-J_{S}i\mbox{\boldmath{$\xi$}}_{f_{i}}\cdot {\mbox{\boldmath{$\tau$}}}
_{\sigma \sigma ^{\prime }} & J_{C}\Phi _{i}^{\ast }\tau _{\sigma \sigma
^{\prime }}^{0} \\ 
J_{C}\Phi _{i}\tau _{\sigma \sigma ^{\prime }}^{0} & \sum\limits_{\alpha
=x,y,z}(-J_{S}i\xi _{c_{i}}^{\alpha }-B\delta _{\alpha z}).\tau _{\sigma
\sigma ^{\prime }}^{\alpha }
\end{array}
\right) \left( 
\begin{array}{c}
c_{i\sigma ^{\prime }} \\ 
f_{i\sigma ^{\prime }}
\end{array}
\right)& \nonumber \\
&+J_{C}\sum_{i}\Phi _{i}^{\ast }\Phi _{i}+J_{S}\sum_{i}
\mbox{\boldmath{$\xi$}}_{f_{i}}.\mbox{\boldmath{$\xi$}}_{c_{i}}.&
\end{eqnarray}
Introducing the change of variables $\xi _{c_{i}}^{\alpha }=\xi
_{c_{i}}^{\alpha }-iB/J_{S}$, we connect the $f$- magnetization and the $ff$
- dynamical spin susceptibility to the Hubbard Stratonovich field $
\mbox{\boldmath{$\xi$}}_{f_{i}}$ 
\[
M_{f}^{z}=-\frac{1}{\beta }\frac{\partial LnZ}{\partial B_{z}}=i\left\langle
\xi _{f_{i}}^{z}\right\rangle 
\]
\begin{equation}
\label{11}
\chi _{ff}^{\alpha \beta }=-\frac{1}{\beta }\frac{\partial ^{2}LnZ}{\partial
B^{\alpha }\partial B^{\beta }}=-\left\langle \xi _{f_{i}}^{\alpha }\xi
_{f_{i}}^{\beta }\right\rangle +\left\langle \xi _{f_{i}}^{\alpha
}\right\rangle \left\langle \xi _{f_{i}}^{\beta }\right\rangle.
\end{equation}
Using the expression (\ref{7}) for the boson propagator ${\bf D}_{S}^{-1}(
{\bf q})$, we get for the Matsubara spin susceptibility 
\begin{equation}
\label{12}
\chi _{ff}({\bf {q}},i\omega _{\nu })=\frac{\left\langle \varphi _{ff}({\bf {
q}},i\omega _{\nu })\right\rangle }{1-J_{S}^{2}[\left\langle \varphi _{ff}(
{\bf {q}},i\omega _{\nu })\right\rangle \left\langle \varphi _{cc}({\bf {q}}
,i\omega _{\nu })\right\rangle -\left\langle \varphi _{fc}({\bf {q}},i\omega
_{\nu })\right\rangle ^{2}-\frac{2}{J_{S}}\left\langle \varphi _{fc}({\bf {q}
},i\omega _{\nu })\right\rangle ]}.
\end{equation}

The diagrammatic representation of Equation (\ref{12}) is represented in Fig. 2 
when the averaging is omitted.
In the low frequency limit, one can easily check that the dynamical spin
susceptibility may be written in terms of intra- and inter-band
suceptibilities corresponding respectively to particle-hole pair excitations
within the lower $\alpha $ band and from the lower $\alpha $ to the upper $
\beta $ band.

\begin{equation}
\label{13}
\chi _{ff}({\bf {q}},i\omega _{\nu })=\frac{\left\langle \overline{\chi }
_{\alpha \alpha }({\bf {q}},i\omega _{\nu })\right\rangle +\left\langle 
\overline{\chi }_{\alpha \beta }({\bf {q}},i\omega _{\nu })\right\rangle }{
1-J_{S}^{2}\left\langle \chi _{\alpha \alpha }({\bf {q}},i\omega _{\nu
})\right\rangle \left\langle \overline{\chi }_{\alpha \beta }({\bf {q}}
,i\omega _{\nu })\right\rangle }.
\end{equation}

The expressions of the susceptibilities $\chi _{\alpha \alpha }({\bf {q}}
,i\omega _{\nu })$ and $\overline{\chi }_{\alpha \beta }({\bf {q}},i\omega
_{\nu })$ corresponding to intra- and inter-band particle-hole excitations
are given in Appendix B.  The latter expression is
reminiscent of the behaviour proposed by Bernhoeft and Lonzarich \cite{bernhoeft} to explain
the neutron scattering observed in $UPt_{3}$ with the existence of both a "slow" 
and a "fast" component in $\chi^{"}(\bf{q},\omega)/{\omega}$ due to spin-orbit 
coupling. Also in a phenomenological way, the same type of feature has been suggested 
in the duality model developed by Kuramoto and Miyake \cite{kuramoto90}. To our knowledge, the proposed approach
provides the first microscopic derivation from the Kondo lattice model of
such a behaviour.
Expanding the various susceptibilities $\varphi (
{\bf {q}},i\omega _{\nu },r_{0},\lambda _{0},\xi _{f_{i}},\xi _{c_{0}})$ up
to the second order in $\xi _{f_{i}}$, and making use of Eq. (\ref{9}), one can
draw from Eq. (\ref{13}) the following expression of the dynamical spin
susceptibility (taking the analytical continuation $i\omega _{\nu
}\rightarrow \omega +i\delta )$ 
\begin{equation}
\label{14}
\chi _{ff}({\bf {q}},\omega )=\frac{\chi _{\alpha \alpha }({\bf {q}},\omega
)+\overline{\chi }_{\alpha \beta }({\bf {q}},\omega )}{1-J_{S}^{2}\chi
_{\alpha \alpha }({\bf {q}},\omega )\overline{\chi }_{\alpha \beta }({\bf {q}
},\omega )+\lambda \sigma _{f}^{2}}.
\end{equation}
where $\lambda$ is a constant which can be evaluated from the expansion of 
$\left\langle \chi _{\alpha \alpha }({\bf {q}},i\omega _{\nu
})\right\rangle$ and $\left\langle \overline{\chi }_{\alpha \beta }({\bf {q}},i\omega _{\nu })\right\rangle$
up to the second order in $\sigma _{f}^{2}$. 
The variance $\sigma _{f}^{2}$ defined by Eq. (\ref{8}) can as well be expressed as
a function of $\chi _{ff}({\bf {q}},i\omega _{\nu })$ 
\begin{equation}
\label{15}
\sigma _{f}^{2}=\sum_{{\bf {q}},i\omega _{\nu }}\chi _{ff}{({\bf {q}}
,i\omega _{\nu })}=\frac{3}{\pi }\sum_{{\bf {q}}}\int_{0}^{+\infty }\coth 
\frac{\beta \omega }{2}\chi _{ff}^{"}{({\bf {q}},\omega )d\omega }.
\end{equation}

The two last equations Eqs. (\ref{14}-\ref{15}) provide a self-consistent determination of $
\sigma _{f}^{2}$ and hence of the spin susceptibility. We will successively
present now (i) in section IV, the result for the dynamical spin
susceptibility in the random phase approximation (RPA) giving information on
the nature of the collective mode that is involved; (ii) then in section V,
the implications of that mode in the critical phenomena around the quantum
critical point.

\section{RPA-dynamical spin susceptibility near the antiferromagnetic wave
vector : discussion on the nature of the collective mode}

As pointed out at the end of section II, the random phase approximation
(RPA) corresponds to the absence of any averaging in the different
equations. Thereby the expression of the dynamical spin susceptibility at
the RPA level is given by Eq. (\ref{14}) taking $\sigma _{f}^{2}=0$. We now discuss
the ${\bf {q}}$- and $\omega $- dependence of the RPA dynamical spin
susceptibility around the antiferromagnetic wave-vector ${\bf Q}$.

The bare intraband susceptibility $\chi _{_{\alpha \alpha }}({\bf {Q}}+{\bf {
q}},\omega )$ is well approximated\ at $\left| {\bf {q}}\right| \ll \left| 
{\bf {Q}}\right| $ by a lorentzian $\chi _{\alpha \alpha }({\bf {Q}}+{\bf {q}},
\omega )=\rho _{\alpha\alpha}/(1-i\omega/{\Gamma _{0}}+bq^{2})$ where $\rho _{\alpha \alpha }=
\chi _{\alpha \alpha }^{^{\prime }}({\bf {Q}},0)$ and $\Gamma _{0}$ 
is the relaxation rate of order $\left| y_{F}\right|
=T_{K}$. As expected for the antiferromagnetic case, the relaxation rate
remains finite when $q$ goes to zero reflecting the fact that the
fluctuations of the order parameter are not conserved. The bare interband
susceptibility $\overline{\chi }_{\alpha \beta }
({\bf {Q}}+{\bf {q}},\omega)$ can be considered as purely real and frequency
independent equal to $\rho _{\alpha \beta }$ in the low frequency limit. Fig.3 gives the continuum of the intraband and interband particle-hole
excitations $\chi_{\alpha \alpha }^{\prime\prime}\neq 0$ and $\overline{\chi}
_{\alpha \beta }^{\prime\prime}\neq 0$. Due to the presence of the
hybridization gap in the density of states, the latter continuum displays a
gap ranging from $2\sigma _{0}$, the value of the direct gap for the wave-vector 
${\bf {0}}$, to $2\left| y_{F}\right| $, the value of the indirect gap at $
{\bf {Q}}$. The RPA dynamical spin susceptibility is given by

\begin{equation}
\label{18}
\chi _{ff}^{_{RPA}\text{ }\prime \prime }({\bf {Q}}+{\bf {q}},\omega
)=\omega \frac{\chi _{ff}^{\prime }({\bf {Q}}+{\bf {q}})\Gamma (q)}{{\omega}
^{2}+{\Gamma (q)}^{2}}
\end{equation}
with

\[
\chi _{ff}^{\prime }({\bf {Q}}+{\bf {q}})=\frac{\rho _{\alpha
\alpha }+\rho _{\alpha \beta }}{(1-I_{q})}
\]

\[
\Gamma (q)=\Gamma _{0}(1-I_q)
\]
where $I_{q}=I-bq^{2}$ and $I=J_{S}^{2}\rho _{\alpha \alpha }\rho _{\alpha \beta }$ 
(of order $1$ near the antiferromagnetic instability).

The bulk staggered susceptibility $\chi _{ff}^{^{\prime }}({\bf {Q}})$ diverges
at $I=1$. The frequency dependence of $\chi _{ff}^{_{RPA}\text{ }\prime \prime }
({\bf {Q}}+{\bf {q}},\omega)$ is a lorentzian with a vanishing relaxation 
rate $\Gamma (0)$ at the antiferromagnetic transition. The corresponding excitation
can be analyzed as an antiferromagnetic paramagnon mode that softens at the magnetic
transition. Eq. (\ref{18}) provides us with the dispersion of that mode. The corresponding
dynamical exponent $z$ is found to be equal to $2$ due to the overdamping of
the mode when enters the continuum of the $\alpha $-$\alpha $ particle-hole
excitations. We will show how the dynamic exponent $z$ strongly affects the
static critical behavior. This is due to the fact that, contrary to the $T>0$
phase transitions, statics and dynamics are intimately linked in quantum ($
T=0$) phase transitions. We turn now to
the resulting quantum-classical crossover occuring at finite temperature.

\section{Critical phenomena around the antiferromagnetic quantum critical
point (AF-QCP)}

The parameter which controls the temperature dependence of the
thermodynamical variables near the AF-QCP is $\sigma _{f}^{2}$ which
expresses the thermal local fluctuations of the staggered magnetization.
Using Eqs. (\ref{15}) and (\ref{18})

\begin{equation}
\label{19}
\sigma _{f}^{2}=\sum_{{\bf q}}S({\bf {Q}}+{\bf {q}})\sim \int S({\bf {Q}}+
{\bf {q}})q^{d-1}dq,
\end{equation}
where $S({\bf {Q}}+{\bf {q}})$ is the static form factor 
\begin{equation}
\label{20}
S({\bf {Q}}+{\bf {q}})=\frac{3}{\pi }\int_{0}^{\infty }\coth (\beta \omega
/2)\chi _{ff}^{\text{ }\prime \prime }({\bf {Q}}+{\bf {q}},\omega )d\omega.
\end{equation}

Depending on the temperature-scale considered, quantum or classical behavior
is observed. In general, the thermal fluctuations $\sigma _{f}^{2}$ of the
magnetization should depend on the total self-consistent renormalized
dynamical spin susceptibility $\chi _{ff}"({\bf {Q}}+{\bf 
{q}},\omega )$ as defined in Eq.(\ref{13}). However, in practice, we will content
with its truncated form $\chi _{ff}^{_{RPA}\text{ }\prime \prime }({\bf {Q}}+
{\bf {q}},\omega )$ derived at the RPA level. The
temperature dependence of the static form factor $S({\bf {Q}}+{\bf {q}})$ is
given by

\begin{equation}
\label{21}
S({\bf {Q}}+{\bf {q}})=\frac{6T}{\pi }\chi
_{ff}^{\prime }({\bf {Q}}+{\bf {q}}) \tan ^{-1}(\frac{T}{\Gamma (q)}).
\end{equation}
A first cross-over temperature $T_{I}=2\Gamma_{0} (1-I)$ appears which separates the quantum from the
classical regime

\bigskip

(i) $T<T_{I}$ : quantum regime. One can show that the
static form factor $S({\bf {Q}}+{\bf {q}})$ exhibits a quadratic temperature
dependence

\begin{equation}
\label{22}
S({\bf {Q}}+{\bf {q}})=\frac{6}{\pi}\frac{\chi _{ff}^{\prime }({\bf {Q+q}})}{\Gamma
_{0}(1-I_{q})}T^{2},
\end{equation}
The thermal local fluctuations ${\bf \sigma }
_{f}{}^{2}$ of the magnetization are also quadratic as a function of the
temperature. They can be easily evaluated from Eq.(\ref{21}). Taking advantage
of the fact that the integrand $S({\bf {Q}}+{\bf {q}})q^{d-1}$ is peaked at $
q_{1}=\sqrt{(1-I)/b}$, the temperature dependence of ${\bf \sigma }_{f}{}^{2}$ is
also found to be quadratic 
\begin{equation}
\label{23}
{\bf \sigma }_{f}{}^{2}=\frac{3}{2\pi}\frac{\chi _{ff}^{\prime }({\bf {Q}})}{
\Gamma _{0}(1-I)}T^{2},
\end{equation}
The physics is quantum in the sense that
the fluctuations on the scale of the magnetic correlation length $\xi $ have
energy much greater that the temperature $k_{B}T$.

\bigskip

(ii) $T>T_{I}$ : classical regime. In this regime, the static form factor 
$S({\bf {Q}}+{\bf {q}})$ shows a linear temperature dependence. As long as $q$
is smaller than the thermal cut-off $q^{\ast }$ defined by $\omega _{\max
}(q^{\ast })=T$, $S({\bf {Q}}+{\bf {q}})$ takes the following form

\begin{equation}
\label{24}
S({\bf {Q}}+{\bf {q}})=3\chi _{ff}^{\prime }({\bf {Q+q}})T.
\end{equation}
The thermal local fluctuations $\sigma _{f}{}^{2}$ of the
magnetization are deduced from Eq. (\ref{21}). The main contribution arises from the
integration over $\left[ q_{1},q^{\ast }\right] $. The result depends on the
dimensionality $d$. One finds $\sigma_{f}{}^{2}\sim T^{3/2}$ at $d=3$
and $\sigma_{f}^{2}\sim T\ln T$ at $d=2$. This classical regime
corresponds to the case where the energy of the mode on the scale of $\xi $
becomes less than $k_{B}T$.

Table I recapitulates the expressions that we get for the thermal local
fluctuations $\sigma _{f}^{2}$ of the staggered magnetization
depending on the temperature scale and on the dimensionality. Following Eq. (\ref{14}), the staggered spin susceptibility is given by

\begin{equation}
\chi _{Q}^{^{\prime }}=\frac{\chi _{ff}^{0^{\prime }}(Q)}{1-I+\lambda \sigma
_{f}^{2}},
\end{equation}
where $\chi _{ff}^{0^{\prime }}(Q)=\rho _{\alpha \alpha }+\rho _{\alpha
\beta }$. Incorporating
the values of the thermal local fluctuations ${\bf \sigma }_{f}{}^{2}$
reported in Table I, we are able to identify a second cross-over temperature 
$T_{II\text{ }}$. Above $T_{II\text{ }}$, ${\bf \sigma }_{f}{}^{2}$ becomes
larger than $(1-I)$ and all the physical quantities are controlled by the
temperature only. $T_{II\text{ }}\sim (1-I)^{2/3}$ at $d=3$ and $T_{II\text{ 
}}\sim (1-I)$ at $d=2$. As expected in the two-dimensional case, the $T=0$
transition is at its upper critical dimension since $z+d=4$ in the
antiferromagnetic case characterized by a dynamic exponent $z=2$. That
corresponds to the marginal case for which the intermediate regime $
T_{I}<T<T_{II}$ is squeezed out of existence. The results for the $d=3$ and $
d=2$ cases are respectively summarized in Figures 4 and 5 which picture the
different regimes of behaviors reached depending on the values of the
temperature $T$ and of the control parameter $I=J^{2}\rho _{\alpha \alpha
}\rho _{\alpha \beta }$. When $I>1$, long-range antiferromagnetic order
occurs when $T$ is smaller than the N\'{e}el temperature $T_{N}$. 
$T_{N}\sim (I-1)^{2/3}$ at $d=3$ and $T_{N}\sim (I-1)$ at $d=2$.

\bigskip
\section{Discussion}

The low temperature physics is dominated by the presence of an antiferromagnetic
paramagnon mode that softens at the magnetic transition.
The results depend crucially on the value of $d+z$ where $d$ is the spatial
dimension and $z$ is the dynamic exponent associated to that mode. Due to
the overdamping of the mode when it enters the continuum of intraband
particle-hole excitations, the dynamic exponent $z$ is found to be equal to $
2$ in the antiferromagnetic case. As far as the dimensionality is concerned,
there is no doubt about the dimensionality associated with the heavy
quasiparticles in those systems that is clearly of 3. However the question
about the dimensionality of the critical magnetic fluctuations which the
quasiparticles are coupled to, is still an open problem. Up to now, it has
been assumed that the critical fluctuations of the QPT are dominated by the
three-dimensional antiferromagnetic correlations. This would impose a
description by a quantum critical theory with $d=3$, $z=2$. Recently Rosch
et al \cite{rosch} proposed, on the basis of the neutron scattering data in $
CeCu_{6-x}Au_{x}$, that the critical magnetic fluctuations are
two-dimensional which leads to a QCP with $d=2$, $z=2$. Let us summarize the
different regimes of behaviors that we get in each of those two cases ($z=2$
, $d=3$ or $2$) depending on the values of the temperature $T$ and of the
control parameter $I=J^{2}\rho _{\alpha \alpha }\rho _{\alpha \beta }$. As one 
can see, the results that we obtain are very similar to those established by 
Hertz \cite{hertz} and Millis \cite{millis93} using renormalization group approaches in the spin-fluctuation 
theory.
\bigskip

$(i)$ Case $d=3$. A long-range antiferromagnetic phase occurs when $I>1$
below the N\'{e}el temperature $T_{N}\sim (I-1)^{2/3}$. A first cross-over
temperature $T_{I}\sim (1-I)$ separates the quantum from the classical
regime. In the regime $I$ $(I<1)$ and $T<T_{I})$, the physics is quantum
in the sense that the energy of the relevant mode on the scale of $\xi $ is
greater than the temperature $k_{B}T$. Since $d+z>4$, the $T=0$ phase
transition is above its upper critical dimension and the various physical
quantities depend upon the parameter $I$. The magnetic correlation length
diverges at the magnetic transition $\xi \sim 1/\sqrt{(1-I)}$ and the
staggered spin susceptibility behaves as $\chi _{Q}^{^{\prime }}=\chi
_{ff}^{0^{\prime }}(Q)/(1-I+aT^{2})$. Regime $II$ and $III$ are both
classical regimes characterized by large thermal effects since the
fluctuations on the scale of $\xi $ have energy much smaller than $k_{B}T$.
In Regime $II$ $(T_{I}<T<T_{II}$ with $T_{II}\sim (1-I)^{2/3})$, the
magnetic correlation length $\xi \sim 1/\sqrt{(1-I)}$ is still controlled by 
$(1-I)$ even though modes at the scale of $\xi $ have energies less than $
k_{B}T$. On the contrary, the staggered spin susceptibility is sensitive to
the thermal fluctuations $\chi _{Q}^{^{\prime }}=\chi _{ff}^{0^{\prime
}}(Q)/(1-I+aT^{3/2})$. In Regime $III$ $(T>T_{II})$, the thermal dependence
of physical quantities becomes universal and both $\xi $ and $\chi
_{Q}^{^{\prime }}$ are controlled by the temperature : $\xi \sim 1/T^{3/4}$
and $\chi _{Q}^{^{\prime }}=\chi _{ff}^{0^{\prime }}(Q)/T^{3/2}.$

\bigskip

$(ii)$ Case $d=2$. Then, since $z=2$, $z+d=4$ and the $T=0$ phase transition
is at its upper critical dimension. The physics is qualitatively similar to
the case $d=3$ with stronger fluctuation effects particularly in the
classical regime. A long-range antiferromagnetic phase occurs when $I>1$
below the N\'{e}el temperature $T_{N}\sim (I-1)$. For $d=2$, the two
cross-over temperatures $T_{I}$ and $T_{II}$ coincide so that Regime $II$ of
Figure 4 is squeezed out of existence. Regime $I$ ($T<T_{I}$ with $T_{I}\sim
(1-I))$ is the quantum regime in which the thermal fluctuations are
negligible : $\xi \sim 1/\sqrt{(1-I)}$ and $\chi _{Q}^{^{\prime }}=\chi
_{ff}^{0^{\prime }}(Q)/(1-I+aT^{2})$. Regime $III$ $(T>T_{I})$ is the unique
classical regime characterized by very strong fluctuation effects since $
k_{B}T$ is larger than the energy of the relevant mode on the scale of $\xi $
. Both the magnetic correlation length and the staggered spin susceptibility
are then controlled by the temperature : $\xi \sim 1/\sqrt{T\ln T}$ and $
\chi _{Q}^{^{\prime }}=\chi _{ff}^{0^{\prime }}(Q)/(T\ln T)$.

\section{Concluding remarks}

\bigskip

We considered the $S=1/2$ Kondo lattice model in a self-consistent one-loop
approximation starting from a generalized Hubbard-Stratonovich
transformation of the Kondo interaction term. The model exhibits a zero
temperature magnetic phase transition at a critical value of the Kondo
coupling. The transition is usually antiferromagnetic but it may be
incommensurate depending on the bandstructure considered. The
low temperature physics is controlled by a collective mode that softens at the
antiferromagnetic transition with a dynamic exponent $z$ equal to $2$. 

A quantum-classical crossover occurs at a temperature $T_{I}$ related to
the characteristic energy-scale of that mode. Heavy-fermion systems are
usually believed to be $3$-dimensional. However since some recent inelastic
neutron scattering experiments performed in $CeCu_{6-x}Au_{x}$ show that the
critical magnetic fluctuations which the quasiparticles are coupled to, are $
2$-dimensional, we both considered the cases $d=2$ and $d=3$ with $z=2$. The
low-temperature behavior of the system is deduced, with predictions for the
temperature dependence of the physical quantities such as the magnetic
correlation length $\xi $ and the staggered susceptibility $\chi
_{Q}^{\prime }$. A second crossover temperature $T_{II}$ appears above which
the thermal dependence of $\xi $ and $\chi _{Q}^{\prime }$ becomes universal
and is uniquely controlled by the temperature. It would be very interesting
now to study the temperature behavior of other physical quantities as the
specific heat or the transport properties. 

A number of significant issues remain to be addressed. Among others, we will 
mention the possibility of getting a different value of the dynamic exponent 
for instance $z=1$ if nesting effects are considered corresponding to an 
absence of damping of the mode when located in the gap of excitations around 
the antiferromagnetic vector. Disorder is expected to play a crucial role 
in this problem. In that direction, some recent works \cite{hlubina,rosch99} 
pointed out the formation of some "hot lines", i.e., points at the Fermi 
surface linked by the magnetic {\bf Q} vector. On the hot lines, the 
quasiparticle scattering rate is linear in temperature while, away from 
the hot lines, it acquires the standard Fermi liquid form in $T^2$.  
At low temperature, the "cold" regions are shown to short-circuit the 
scattering near those lines, and induce a $T^2$ behavior of the resistivity. 
The disorder is found to emphasize the contribution of the hot lines, 
eventually leading to a non-Fermi liquid behavior. It would be worthy 
to study the effects of the "hot lines" in presence of disorder within 
the critical phenomena description presented in this paper.

\vspace{0.2in}

We would like to thank Nick Bernhoeft, Piers Coleman, Jacques Flouquet, Patrick A. Lee 
and Gilbert Lonzarich for very useful discussions on this work.

\vspace{0.2in}

$^*$  Also at the Centre National de la Recherche Scientifique (CNRS)

\vfill\eject

\subsection{Appendix A: Saddle-Point}

The saddle-point solution is obtained from Eq. (\ref{4}) for space and time
independent fields $\Phi _{0}$, $\lambda _{0}$, $\xi _{f_{0}}$ and $\xi
_{c_{0}}$. In the magnetically-disordered regime ($\xi _{f_{0}}=\xi
_{c_{0}}=0)$, it leads to renormalized bands $\alpha $ and $\beta $ as
schematized in Figure 1. Noting $\sigma _{0}^{(\ast )}=J_{C}\Phi _{0}^{(\ast
)}$ and $\varepsilon _{f}=\lambda _{0}$, $\alpha _{k\sigma }^{\dagger }|0>$
and $\beta _{k\sigma }^{\dagger }|0>$ are the eigenstates of

\begin{equation}  \label{eq7}
{\bf G}_{0}^{-1 \sigma }({\bf {k},\tau )=\left( 
\begin{array}{cc}
\partial _{\tau }+\varepsilon _{k} & \sigma _{0}^{\ast} \\ 
\sigma _{0} & \partial _{\tau }+\varepsilon _{f}
\end{array}
\right)},
\end{equation}
with respectively the eigenenergies $\left( \partial _{\tau
}+E_{k}^{-}\right)$ and $\left( \partial _{\tau }+E_{k}^{+}\right) $. In the
notations: $x_{k}=\varepsilon _{k}-\varepsilon _{f}$, $y_{k}^{\pm
}=E_{k}^{\pm }-\varepsilon _{f}$ and $\Delta _{k}=\sqrt{x_{k}^{2}+4\sigma
_{0}^{2}}$, we get

\begin{equation}
y_{k}^{\pm }=\left( x_{k}\pm \Delta _{k}\right) /2.
\end{equation}
Let us note $U_{k\sigma }^{\dagger }$ the matrix transforming the initial
basis $(c_{k\sigma }^{\dagger }$ $f_{k\sigma }^{\dagger })$ to the
eigenbasis $(\alpha _{k\sigma }^{\dagger }$ $\beta _{k\sigma }^{\dagger })$.
The Hamiltonian being hermitian, the matrix $U_{k\sigma }$ is unitary : $
U_{k\sigma }U_{k\sigma }^{\dagger }=U_{k\sigma }^{\dagger }U_{k\sigma }=1$.
In the notation $U_{k\sigma }^{\dagger }=\left( 
\begin{array}{cc}
-v_{k} & u_{k} \\ 
u_{k} & v_{k}
\end{array}
\right) $, we have 
\begin{equation}
u_{k}=\frac{-\sigma _{0}/y_{k}^{-}}{\sqrt{1+(\sigma _{0}/y_{k}^{-})^{2}}}=
\frac{1}{2}\left[ 1+\frac{x_{k}}{\Delta _{k}}\right]   \label{eq9}
\end{equation}
\[
v_{k}=\frac{1}{\sqrt{1+(\sigma _{0}/y_{k}^{-})^{2}}}=\frac{1}{2}\left[ 1-
\frac{x_{k}}{\Delta _{k}}\right].
\]

The saddle-point equations together with the conservation of the number of
conduction electrons are written as 
\begin{equation}
\sigma _{0}=\frac{1}{N_{S}}J_{C}\sum_{k\sigma }u_{k}v_{k}~n_{F}(E_{k}^{-})
\label{eq10}
\end{equation}
\[
1=\frac{1}{N_{S}}\sum_{k\sigma }u_{k}^{2}~n_{F}(E_{k}^{-})
\]
\[
n_{c}=\frac{1}{N_{S}}\sum_{k\sigma }v_{k}^{2}~n_{F}(E_{k}^{-}).
\]
Their resolution at zero temperature and for $n_c$ close to $1$ leads to 
\begin{equation}
\left| y_{F}\right| =D\exp \left[ -2/\left( \rho _{0}J_{C}\right) \right] 
\label{eq11}
\end{equation}
\[
2\rho _{0}\sigma _{0}^{2}/\left| y_{F}\right| =1
\]
\[
\mu =0,
\]
where $y_{F}=\mu -\varepsilon _{F}$ and $\rho _{0}$ is the bare density of
states of conduction electrons ($\rho _{0}=1/2D$ for a flat band). Noting $
y=E-\varepsilon _{F}$, the density of states at the energy E is $\rho \left(
E\right) =\rho _{0}\left( 1+\sigma _{0}^{2}/y^{2}\right) $. If $n_{c}<1$,
the chemical potential is located just below the upper edge of the $\alpha $
-band. The system is metallic. The density of states at the Fermi level is
strongly enhanced compared with the bare density of states of conduction electrons
: $\rho (E_{F})/\rho _{0}=(1+\sigma _{0}^{2}/y_{F}^{2})\sim 1/(2\rho
_{0}\left| y_{F}\right| )$. This corresponds to the flat part of the $\alpha 
$-band in Figure 1. It is associated to the formation of a Kondo or
Abrikosov-Suhl resonance pinned at the Fermi level resulting of the Kondo
effect. The low-lying excitations are quasiparticles of large effective mass 
$m^{\ast }$ as observed in heavy-Fermion systems. Also note the presence of
a hybridization gap between the $\alpha $ and the $\beta $ bands. The direct
gap of value $2\sigma _{0}$ is much larger than the indirect gap equal to 2$
\left| y_{F}\right| $. The saddle-point solution transposes to N=2 the
large-N results obtained within the slave-boson mean-field theories (SBMFT).

\subsection{Appendix B : Expressions of the different bubbles}

The expressions of the different bubbles appearing in the expression of the
boson propagators (cf. Eq.\ref{7}) are given here (with i=1, 2, m or ff) 
\begin{equation}
\overline{\varphi }_{i}({\bf {q}},i\omega _{\nu })=\varphi _{i}({\bf {q}}
,i\omega _{\nu })+\varphi _{i}(-{\bf {q}},-i\omega _{\nu })
\end{equation}
\begin{eqnarray*}
\varphi _{1}({\bf {q}},i\omega _{\nu }) &=&-\frac{1}{\beta }\sum_{k\sigma
,i\omega _{n}}G_{cf_{0}}^{\sigma }({\bf k+q},i\omega _{n}+i\omega _{\nu
})G_{ff_{0}}^{\sigma }({\bf {k}},i\omega _{n}) \\
\varphi _{2}({\bf {q}},i\omega _{\nu }) &=&-\frac{1}{\beta }\sum_{k\sigma
,i\omega _{n}}G_{cc_{0}}^{\sigma }({\bf k+q},i\omega _{n}+i\omega _{\nu
})G_{ff_{0}}^{\sigma }({\bf {k}},i\omega _{n}) \\
\varphi _{m}({\bf {q}},i\omega _{\nu }) &=&-\frac{1}{\beta }\sum_{k\sigma
,i\omega _{n}}G_{cf_{0}}^{\sigma }({\bf k+q},i\omega _{n}+i\omega _{\nu
})G_{cf_{0}}^{\sigma }({\bf {k}},i\omega _{n})
\end{eqnarray*}
\begin{eqnarray*}
\varphi _{ff}^{\Vert }({\bf {q}},i\omega _{\nu }) &=&-\frac{1}{\beta }
\sum_{k\sigma ,i\omega _{n}}G_{ff_{0}}^{\sigma }({\bf k+q},i\omega
_{n}+i\omega _{\nu })G_{ff_{0}}^{\sigma }({\bf {k}},i\omega _{n}) \\
\varphi _{cc}^{\Vert }({\bf {q}},i\omega _{\nu }) &=&-\frac{1}{\beta }
\sum_{k\sigma ,i\omega _{n}}G_{cc_{0}}^{\sigma }({\bf k+q},i\omega
_{n}+i\omega _{\nu })G_{cc_{0}}^{\sigma }({\bf {k}},i\omega _{n}) \\
\varphi _{fc}^{\Vert }({\bf {q}},i\omega _{\nu }) &=&-\frac{1}{\beta }
\sum_{k\sigma ,i\omega _{n}}G_{fc_{0}}^{\sigma }({\bf k+q},i\omega
_{n}+i\omega _{\nu })G_{fc_{0}}^{\sigma }({\bf {k}},i\omega _{n})
\end{eqnarray*}
\begin{eqnarray*}
\varphi _{ff}^{\bot }({\bf {q}},i\omega _{\nu }) &=&-\frac{1}{\beta }
\sum_{k\sigma ,i\omega _{n}}G_{ff_{0}}^{\uparrow }({\bf k+q},i\omega
_{n}+i\omega _{\nu })G_{ff_{0}}^{\downarrow }({\bf {k}},i\omega _{n}) \\
\varphi _{cc}^{\bot }({\bf {q}},i\omega _{\nu }) &=&-\frac{1}{\beta }
\sum_{k\sigma ,i\omega _{n}}G_{cc_{0}}^{\uparrow }({\bf k+q},i\omega
_{n}+i\omega _{\nu })G_{cc_{0}}^{\downarrow }({\bf {k}},i\omega _{n}) \\
\varphi _{fc}^{\bot }({\bf {q}},i\omega _{\nu }) &=&-\frac{1}{\beta }
\sum_{k\sigma ,i\omega _{n}}G_{fc_{0}}^{\uparrow }({\bf k+q},i\omega
_{n}+i\omega _{\nu })G_{fc_{0}}^{\downarrow }({\bf {k}},i\omega _{n}),
\end{eqnarray*}
where $G_{cc_{0}}^{\sigma }({\bf {k}},i\omega _{n})$, $G_{ff_{0}}^{\sigma }
({\bf {k}},i\omega _{n})$ and $G_{fc_{0}}^{\sigma }({\bf {k}},i\omega _{n})$
are the Green's functions at the saddle-point level obtained by inversing
the matrix $G_{0}^{\sigma }({\bf {k},\tau )}$ defined in Equation (\ref{5}).

The different bubbles $\varphi _{ff}({\bf {q}},i\omega _{\nu })$, $\varphi
_{cc}({\bf {q}},i\omega _{\nu })$ and $\varphi _{fc}({\bf {q}},i\omega _{\nu
})$ can be expressed as a function of the Green's functions associated with
the eigenoperators $\alpha _{k\sigma }^{\dagger }$ and $\beta _{k\sigma
}^{\dagger }$
\begin{equation}
G_{ff}({\bf {k}},i\omega _{n})=u_{k}^{2}G_{\alpha \alpha }({\bf {k}},i\omega
_{n})+v_{k}^{2}G_{\beta \beta }({\bf {k}},i\omega _{n})
\end{equation}
\[
G_{cc}({\bf {k}},i\omega _{n})=v_{k}^{2}G_{\alpha \alpha }({\bf {k}},i\omega
_{n})+u_{k}^{2}G_{\beta \beta }({\bf {k}},i\omega _{n}) 
\]
\[
G_{cf}({\bf {k}},i\omega _{n})=G_{fc}({\bf {k}},i\omega
_{n})=-u_{k}v_{k}[G_{\alpha \alpha }({\bf {k}},i\omega _{n})-G_{\beta \beta
}({\bf {k}},i\omega _{n})], 
\]

where $G_{\alpha \alpha }({\bf k},i\omega _{n})$ and $G_{\beta \beta }({\bf k
},i\omega _{n})$ are the Green's functions associated to the eigenstates $
\alpha _{k\sigma }^{\dagger }|0>$and $\beta _{k\sigma }^{\dagger }|0>$. In
the low frequency limit, one can easily check that the dynamical spin
susceptibility may be written as

\begin{equation}
\chi _{ff}({\bf {q}},i\omega _{\nu })=\frac{\chi _{\alpha \alpha }({\bf {q}}
,i\omega _{\nu })+\overline{\chi }_{\alpha \beta }({\bf {q}},i\omega _{\nu })
}{1-J_{S}^{2}\chi _{\alpha \alpha }({\bf {q}},i\omega _{\nu })\overline{\chi 
}_{\alpha \beta }({\bf {q}},i\omega _{\nu })},
\end{equation}
for both the longitudinal and the transverse parts. 
\[
\chi _{\alpha \alpha }({\bf {q}},i\omega _{\nu })=\frac{1}{\beta }
\sum\limits_{k}\frac{n_{F}(E_{k}^{-})-n_{F}(E_{k+q}^{-})}{i\omega _{\nu
}-E_{k+q}^{-}+E_{k}^{-}} 
\]
\[
\overline{\chi }_{\alpha \beta }({\bf {q}},i\omega _{\nu })=\frac{1}{\beta }
\sum\limits_{k}(u_{k}^{2}v_{k+q}^{2}+v_{k}^{2}u_{k+q}^{2})\frac{
n_{F}(E_{k}^{-})-n_{F}(E_{k+q}^{+})}{i\omega _{\nu }-E_{k+q}^{+}+E_{k}^{-}}.
\]

\vfill\eject

\vfill\eject

\centerline {\bf TABLE CAPTIONS}

\vspace{0.4in}

Table I : Predictions for thermal local fluctuations ${\bf \sigma }
_{f}{}^{2} $ of the staggered magnetization depending on the temperature $T$
and on the dimensionality $d$. $T_{I}\sim (1-I)$ represents the first
cross-over temperature separating the quantum from the classical regime in
the vicinity of the antiferromagnetic-quantum critical point (AF-QCP). 

\vspace{0.4in}

\begin{center}
\begin{tabular}{|c|c|c|}
\hline
&     quantum regime $T<T_{I}$     &     classical regime $T>T_{I}$     \\
\hline
$d=3$ & $ \sigma _{f}^{2}\sim \chi _{ff}^{\prime }(Q)T^{2}/\Gamma _{0} $ &
$ \sigma _{f}^{2}\sim T^{3/2} $ \\
\hline
$d=2$ & $ \sigma _{f}^{2}\sim \chi _{ff}^{\prime }(Q)T^{2}/\Gamma _{0} $ &
$ \sigma _{f}^{2}\sim T\ln T $ \\
\hline
\end{tabular}
\end{center}

\vfill\eject

\vspace{0.4in}

Figure 1: Energy versus wave-vector k for the two bands $\alpha$ and $\beta$. Note 
the presence of a direct gap of value $2\sigma _{0}$ and of an indirect gap of
value $2\left| y_{F}\right|$.

\vspace{0.8in}

\begin{figure}
\centerline{\psfig{file=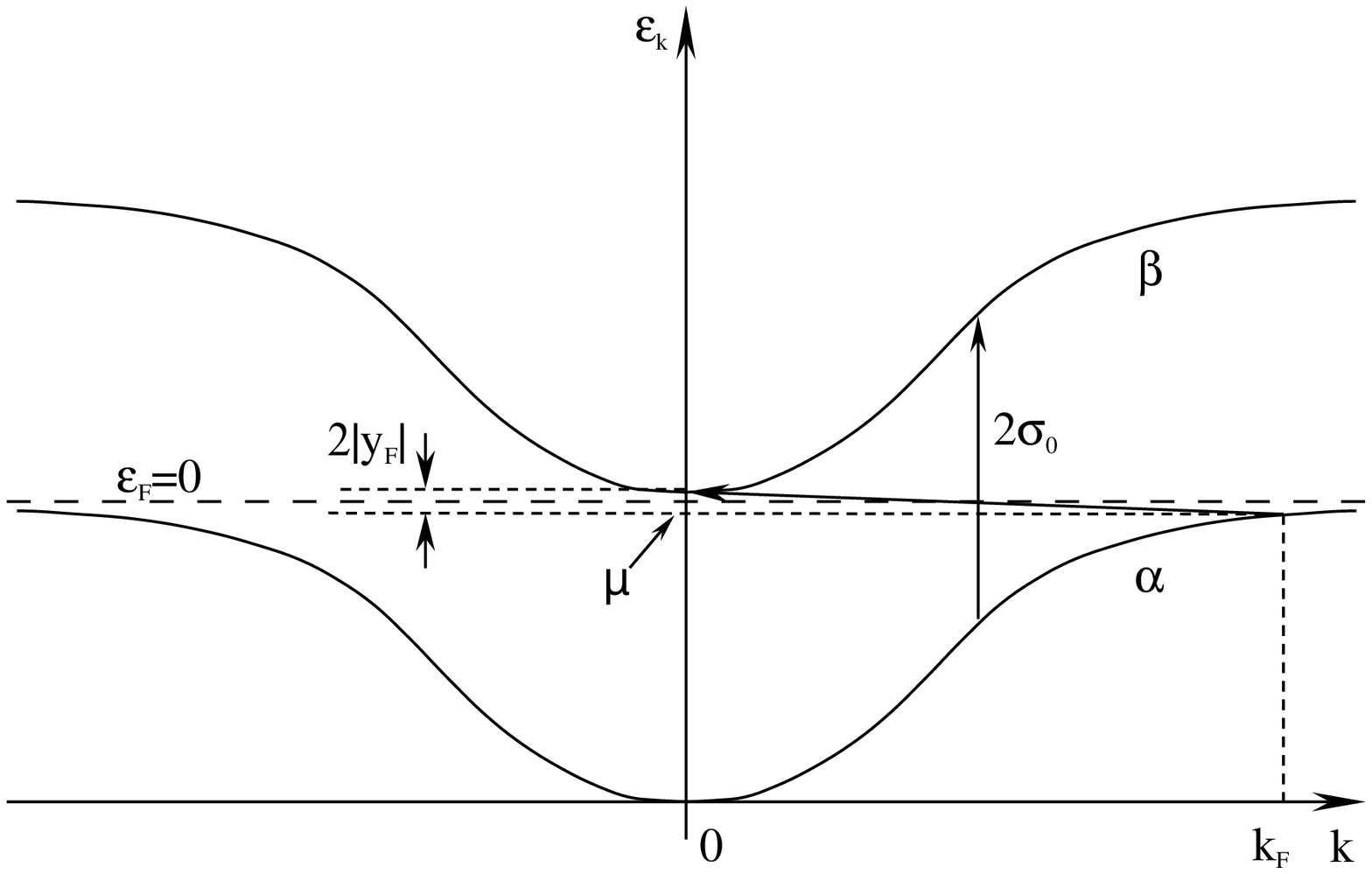,height=9cm,width=16cm}}
\label{fig1}
\end{figure}

\vfill\eject

Figure 2: Diagrammatic representation of Equation (\ref{12}) for the dynamical spin susceptibility $\chi _{ff}(\bf{q},\omega)$.

\vspace{0.8in}

\begin{figure}
\centerline{\psfig{file=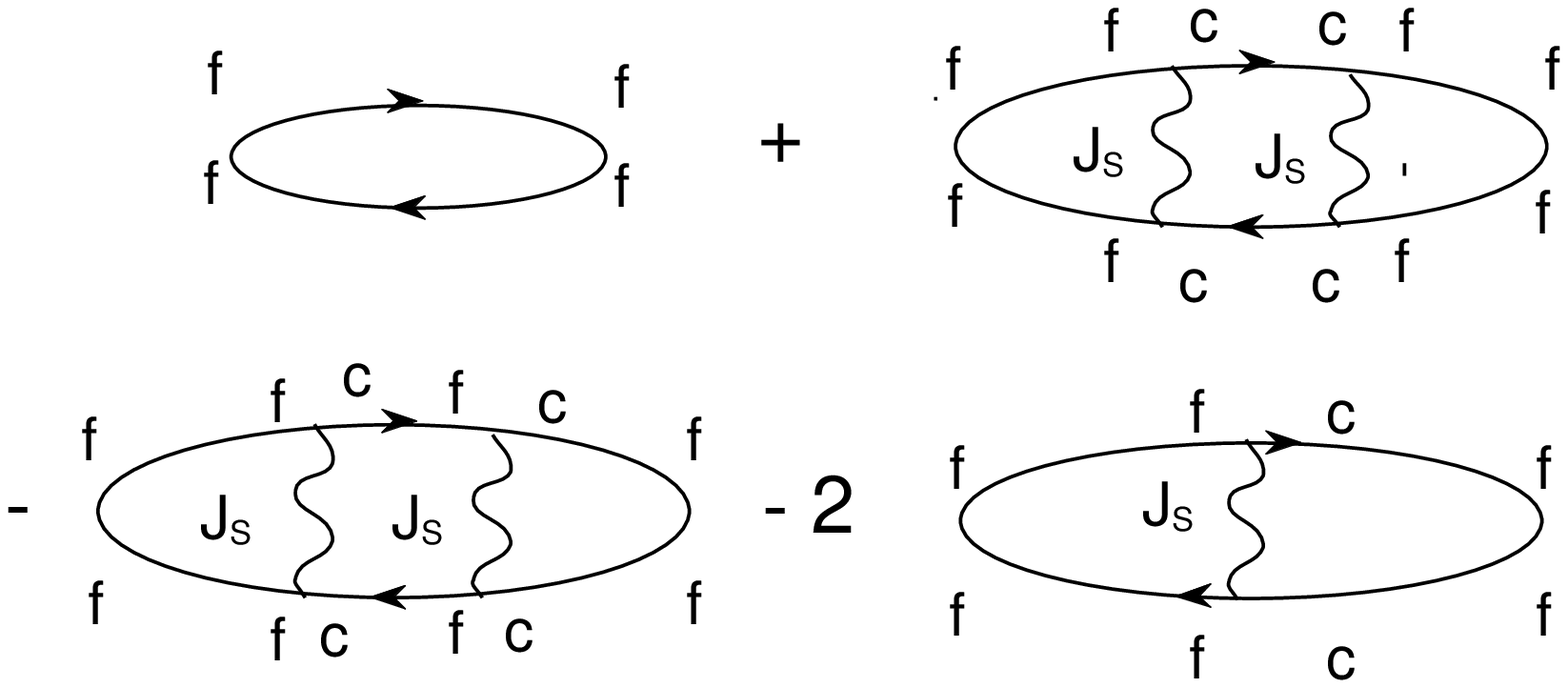,height=9cm,width=16cm}}
\label{fig2}
\end{figure}

\vfill\eject

Figure 3: Continuum of the intra- and interband electron-hole pair excitations 
$\chi _{_{\alpha \alpha}}^{"}(q,\omega)\neq 0$ and $\chi _{_{\alpha \beta}}^{"}(q,\omega)\neq 0$.
Note the presence of a gap in the interband transitions equal to the indirect gap of value 
$2\left| y_{F}\right|$ at $q=k_F$, and to the direct gap of value $2\sigma _{0}$ at $q=0$.
 
\vspace{0.8in}

\begin{figure}
\centerline{\psfig{file=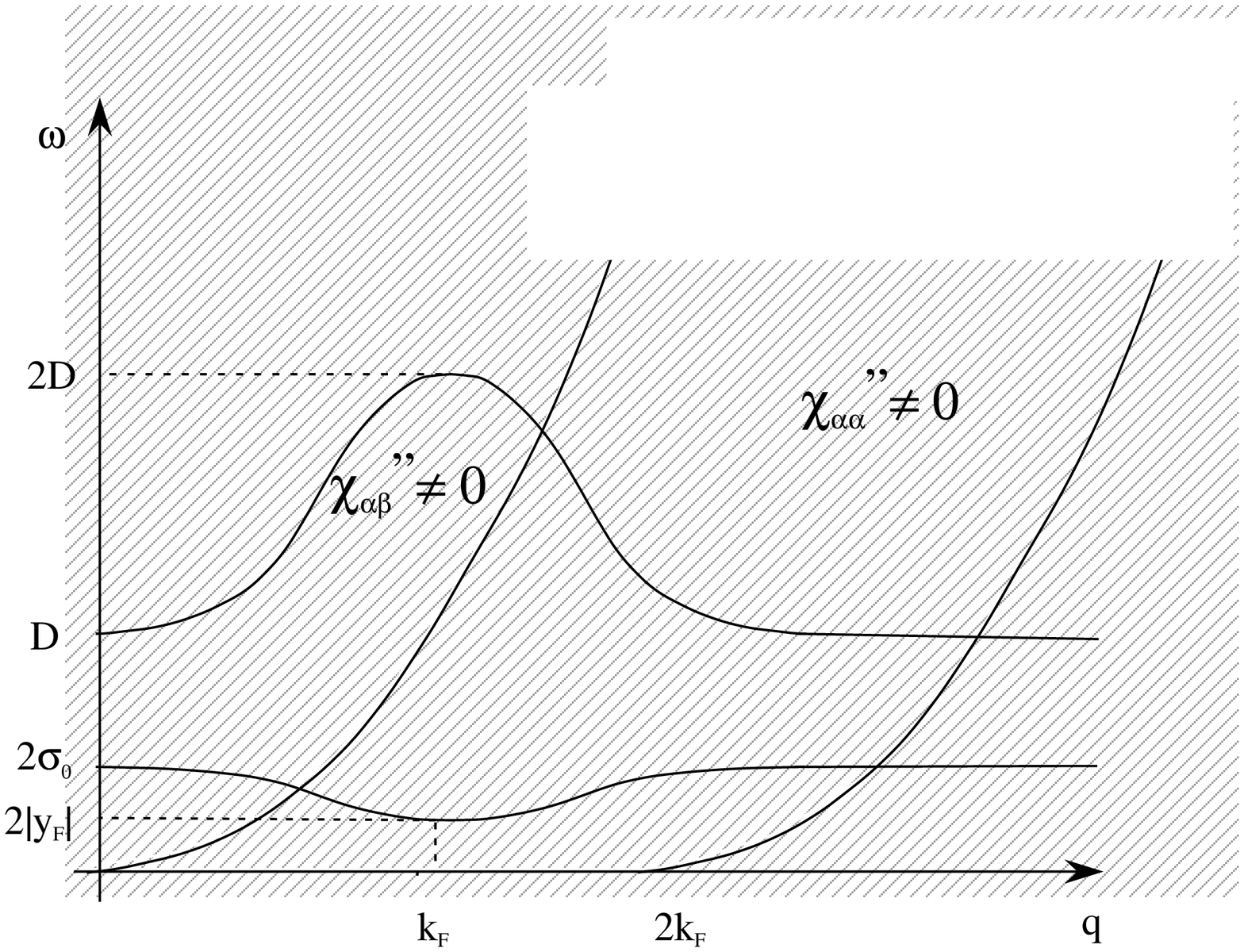,height=9cm,width=16cm}}
\label{fig3}
\end{figure}

\vfill\eject

Figure 4 : Phase diagram in the plane $(T,I)$ for dimension equal to $3$.
The shaded region represents the long-range antiferromagnetic phase bordered
by the N\'{e}el temperature $T_{N}$. The unshaded region marks the
magnetically-disordered regimes $I,$ $II$ and $III$ associated to different
behaviors of the system. Regime $I$ is the quantum regime in which the
energy of the relevant mode on the scale of $\xi $ is much greater than $
k_{B}T$. The magnetic correlation length is $\xi \sim 1/\sqrt{(1-I)}$ and
the staggered spin susceptibility is $\chi _{Q}^{^{\prime }}=\chi
_{ff}^{0^{\prime }}(Q)/(1-I+aT^{2})$. Regime $II$ and $III$ are both
classical regimes in which the thermal effects are important since the
fluctuations on the scale of $\xi $ have energy much smaller than $k_{B}T$.
In Regime $II$, $\xi \sim 1/\sqrt{(1-I)}$ is still controlled by $(1-I)$ but
the staggered spin susceptibility is sensitive to the thermal fluctuations : 
$\chi _{Q}^{^{\prime }}=\chi _{ff}^{0^{\prime }}(Q)/(1-I+aT^{3/2})$. In
Regime $III$, both $\xi $ and $\chi _{Q}^{^{\prime }}$ are controlled by the
temperature : $\xi \sim 1/T^{3/4}$ and $\chi _{Q}^{^{\prime }}=\chi
_{ff}^{0^{\prime }}(Q)/T^{3/2}.$

\begin{figure}
\centerline{\psfig{file=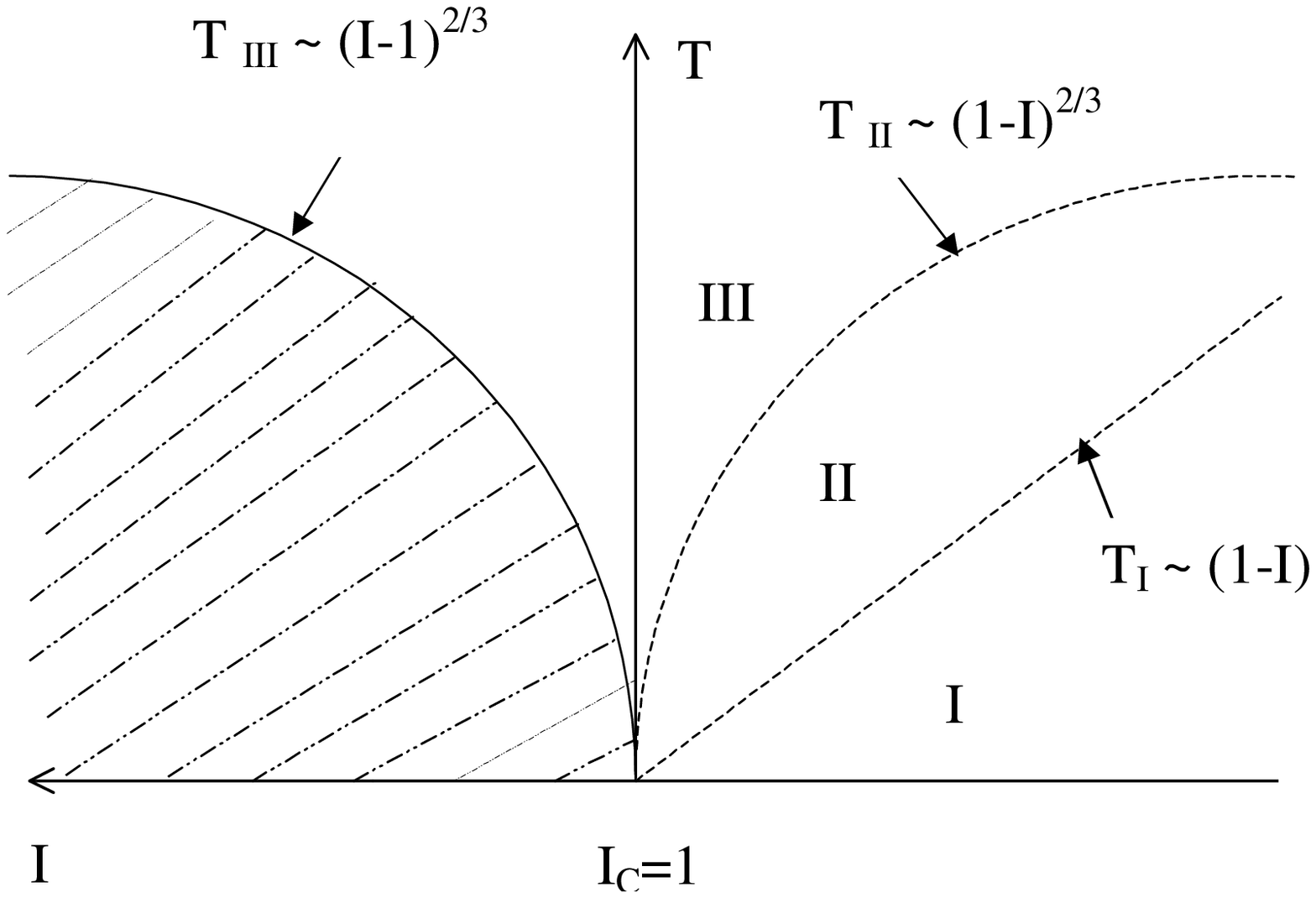,height=16cm,width=16cm}}
\label{fig4}
\end{figure}

\vfill\eject

Figure 5 : Phase diagram in the plane $(T,I)$ for dimension equal to $2$.
The shaded region represents the long-range antiferromagnetic phase bordered
by the N\'{e}el temperature $T_{N}$. The unshaded region marks the
magnetically-disordered regimes $I$ and $III$. In that $d=2$ case, the
equivalent of Regime $II$ in Figure 3 is squeezed out of existence since $
T_{I}=T_{II}$. Regime $I$ is the quantum regime in which the thermal
fluctuations are negligible : $\xi \sim 1/\sqrt{(1-I)}$ and $\chi
_{Q}^{^{\prime }}=\chi _{ff}^{0^{\prime }}(Q)/(1-I+aT^{2})$. Regime $III$ is
the unique classical regime in which both the magnetic correlation length
and the staggered spin susceptibility are controlled by the temperature : $
\xi \sim 1/\sqrt{T\ln T}$ and $\chi _{Q}^{^{\prime }}=\chi _{ff}^{0^{\prime
}}(Q)/(T\ln T)$.

\begin{figure}
\centerline{\psfig{file=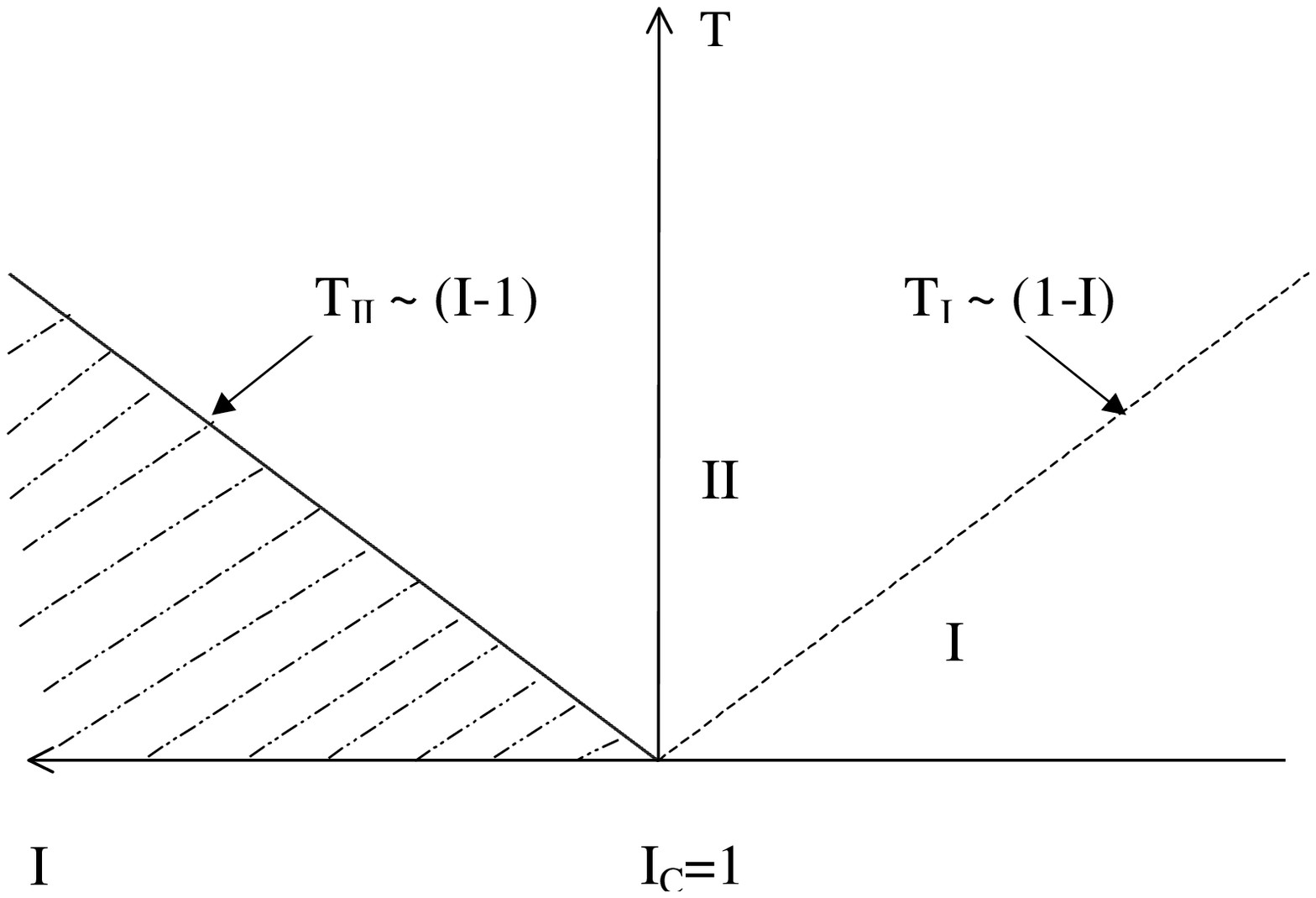,height=16cm,width=16cm}}
\label{fig5}
\end{figure}

\end{document}